\documentclass[aps,pre,twocolumn,groupedaddress,showpacs,floatfix]{revtex4}
\usepackage{graphicx}

\begin{document}

\title{Pair Contact Process with Diffusion: 
       Failure of Master Equation Field Theory}

\author{Hans-Karl Janssen}
\affiliation{Institut f\"ur Theoretische Physik III, 
	Heinrich-Heine-Universit\"at, 40225 D\"usseldorf, Germany}

\author{Fr\'ed\'eric van Wijland} 
\affiliation{Laboratoire de Physique Th\'eorique, Universit\'e de Paris-Sud, 
        F-91405 Orsay cedex, France}

\author{Olivier Deloubri\`ere}
\affiliation{Department of Physics, Virginia Tech, Blacksburg, 
	Virginia 24061-0435, USA}

\author{Uwe C. T\"auber}
\affiliation{Department of Physics, Virginia Tech, Blacksburg, 
	Virginia 24061-0435, USA}

\date{\today}

\begin{abstract}
We demonstrate that the `microscopic' field theory representation, directly
derived from the corresponding master equation, fails to adequately capture
the continuous nonequilibrium phase transition of the Pair Contact Process 
with Diffusion (PCPD). The ensuing renormalization group (RG) flow equations
do not allow for a stable fixed point in the parameter region that is 
accessible by the physical initial conditions. There exists a stable RG fixed 
point outside this regime, but the resulting scaling exponents, in conjunction 
with the predicted particle anticorrelations at the critical point, would be 
in contradiction with the positivity of the equal-time mean-square particle 
number fluctuations. We conclude that a more coarse-grained effective field 
theory approach is required to elucidate the critical properties of the PCPD.
\end{abstract}

\pacs{05.40.-a, 64.60.Ak, 64.60.Ht, 82.20.-w} 

\maketitle


\section{Introduction}
\label{intro}

Phase transitions between different nonequilibrium steady states are frequently
encountered in nature, and determining the associated critical properties is an
important issue. Unfortunately, compared with the situation in thermal
equilibrium, a full classification of nonequilibrium phase transitions is still
in its infancy. We shall focus here on a particular subclass of nonequilibrium 
transitions which separate an `active' phase, characterized by a fluctuating 
order parameter $\phi(\mathbf{r},t)$ with nonzero average
$\langle \phi \rangle$, from an absorbing state wherein 
$\langle \phi \rangle = 0$. In the thermodynamic limit, all degrees of freedom 
remain strictly frozen in such an inactive, absorbing phase \cite{books}.

The universality classes of such transitions are conveniently studied in the
framework of reaction-diffusion processes, even though other descriptions 
abound (surface growth, self-organized criticality) \cite{hinrichsenodorrev}. 
The most prominent representative of absorbing state transitions is the 
{\em contact process} (CP), for under quite generic conditions, namely 
spatially and temporally {\em local} microscopic dynamics, and the absence of 
coupling to other slow fields (thus excluding quenched disorder and the 
presence of conservation laws), active to absorbing state transitions fall into
the CP universality class with scaling exponents that also describe critical 
{\em directed percolation} (DP) clusters \cite{janssen,grassberger}. Yet, the 
very fact that despite considerable effort hardly any experiments have to date 
unambiguously identified the CP/DP critical exponents, hints at the prevalence 
of other universality classes. In simulations, the `{\em parity-conserving}' 
(PC) universality class is also prominent: represented by one-dimensional 
branching and annihilating random walks (BAW) $A \to (m+1) \, A$, 
$A + A \to \emptyset$ with {\em even} $m$, it is characterized by {\em local} 
particle number parity conservation. In contrast, the phase transition in 
low-dimensional BAW with {\em odd} $m$ is governed by DP exponents 
\cite{hinrichsenodorrev,johnuwe}.
 
Novel critical behavior is to be expected when {\em all} involved reactions 
require the presence of neighboring particle pairs 
\cite{grassberger}. The {\em pair contact process with diffusion} (PCPD, or 
{\em annihilation/fission} model \cite{martinuwe}) is conveniently defined 
through the microscopic reaction rules
\begin{equation}
\label{reactionrules}
  A + A \stackrel{\mu}{\to} \emptyset \, , \quad  A + A \stackrel{\mu'}{\to} A
  \, , \quad A + A \stackrel{\sigma}{\to} A + A + A \quad
\end{equation}
(the presence of either pair annihilation $\sim \mu$ or coagulation $\sim \mu'$
suffices), supplemented with particle hopping (diffusion constant $D$) subject 
to mutual exclusion. The latter is crucial for the existence of a well-defined 
active phase and continuous transition. For without restrictions on the 
occupation number per lattice site, the particle density diverges within a 
finite time when $\sigma > \sigma_c = 2 \mu + \mu'$ \cite{martinuwe}. In the 
inactive phase, however, site exclusion should not be relevant. It is then 
easily seen that the absorbing state of the PCPD (as in the PC universality
class \cite{johnuwe}) is governed by the {\em algebraic} density decay of the 
pure pair annihilation process \cite{pelitilee}, {\em viz.} 
$\langle \phi(t) \rangle \sim t^{-1}$ in dimensions $d > 2$, 
$\langle \phi(t) \rangle \sim t^{-d/2}$ for $d < 2$, and 
$\langle \phi(t) \rangle \sim t^{-1} \ln t$ at the (upper) critical dimension 
$d_c = 2$. In contrast, in the CP/DP universality class, the inactive phase is 
characterized by {\em exponential} particle decay and correlations. Recall that
here the branching processes merely require the presence of a single particle:
the third reaction in (\ref{reactionrules}) would simply be replaced with 
$A \to A + A$. Site exclusion is not crucial in this case, as long as the pair 
annihilation or coagulation reactions are included. Alternatively, the
combined first-order reactions $A \to \emptyset$ and $A \to A + A$ {\em with}
site exclusion yield a CP/DP continuous phase transition as well.

Holding the rates $\mu$ and $\mu'$ fixed, there is a critical production rate
$\sigma_c$ at which the transition between the active nonequilibrium steady 
state and the absorbing phase occurs. It is a central issue, in an effort to 
classify nonequilibrium phase transitions, to clarify the precise manner in 
which the particle production mechanism defines the properties of both the 
absorbing state and the universality class of the transition, {\em i.e.}, how 
it affects scaling properties in the vicinity of the critical point. Numerical 
investigations of the PCPD started with Ref.~\cite{carlonhenkelschollwoeck}. It
almost constitutes an euphemism to state that this and the subsequent flurry of
numerical work \cite{hinrichsen1}--\cite{nohpark} have revealed conflicting 
views (see Ref.~\cite{henkelhinrichsen} for a comprehensive overview): For not 
only are the precise numerical values of the critical exponents still being 
debated to this day, but even more striking, the very issue of the PCPD 
universality class has remained controversial. Essentially three scenarios have
been put forward: Either the transition defines a novel independent 
universality class that is yet to be characterized, or it belongs to the CP/DP,
or even to the PC class (the latter perhaps becoming less likely with improving
simulation accuracy). In addition, the emergence of different scaling 
properties depending on the value of the diffusion rate has been claimed.

This inconclusive numerical situation clearly calls for analytical approaches 
to provide further understanding of the elusive continuous phase transition of 
the PCPD with restricted particle occupation numbers. A natural starting point 
is the standard field-theoretic representation of reaction-diffusion systems 
that can be derived directly from the corresponding classical master equation 
\cite{doietal}. Specifically, dynamical renormalization group (RG) studies 
based on such `microscopic' field theories were, {\em e.g.}, successfully 
employed to diffusion-limited annihilation \cite{pelitilee} and even-offspring 
BAW \cite{johnuwe}, as well as the inactive phase of the PCPD without site
occupation restrictions \cite{martinuwe}. In either case, particle
{\em anti}correlations govern the asymptotic scaling regime, as opposed to the
typical clustering behavior in the CP/DP universality class (that includes 
odd-offspring BAW) \cite{janssen}. Here another coarse-graining step takes the 
original `microscopic' master equation representation to Reggeon field theory, 
equivalent to a Langevin equation with `square-root' multiplicative noise, that
serves as the appropriate {\em effective} action for the CP/DP critical 
properties. Thus, one would hope that the continuous nonequilibrium phase 
transition in the PCPD with restricted site occupations should be amenable to 
these powerful tools as well.

Yet it was only recently demonstrated how site exclusions can be consistently 
incorporated into the master equation field theory formalism \cite{fredo}. This
paper reports our study of the ensuing action for the PCPD, constructed in 
Sec.~\ref{fteor}, carefully taking into account the higher-order reactions that
become generated through fluctuations, {\em i.e.}, successive particle 
production processes \cite{martinuwe}. We shall derive and discuss the ensuing 
RG flow equations in Sec.~\ref{rgflo}. We will demonstrate that there exists in
fact {\em no} stable RG fixed point in the physically accessible parameter 
space of the model (wherein all reaction rates are non-negative). Remarkably 
therefore, the microscopic field theory, albeit directly derived from the 
master equation, is {\em not} capable of capturing the PCPD critical point. We 
interpret the appearance of runaway flows as an indication that a crucial 
ingredient was obviously left out when the (naive) continuum limit was taken
\cite{fnote1}. An appropriate effective coarse-grained description might 
require the explicit introduction of separate density fields for the `inert'
solitary random walkers and the clustered particles, respectively, akin to the 
explicit treatment of particle pairs and singlets in Ref.~\cite{hinrichsen2}. 
The RG flow equations do however allow for a fixed point {\em outside} the 
physical region. In Sec.~\ref{crexp} we compute the associated critical 
exponents to second order in an expansion around the upper critical dimension 
$d_c = 2$, and moreover establish exact scaling relations. But we shall see 
that the actual exponent values, when combined with the predicted particle 
anticorrelations at the critical point, violate the positivity of the 
equal-time mean-square particle fluctuations. Hence this RG fixed point is 
clearly unphysical. In conclusion, the construction of a consistent field 
theory description of the PCPD remains an open problem. Similar issues arise 
also for the closely related models that involve solely particle triplet or 
quadruplet reactions \cite{parkhinrichsenkim2,kockelkorenchate1,odor4}.

\section{Master equation field theory representation of the PCPD}
\label{fteor}

The classical master equation kinetics of particles subject to diffusion and 
local `chemical' reactions can be mapped onto a field theory action following 
standard procedures \cite{doietal,pelitilee}. However, for the density to 
remain bounded in the processes (\ref{reactionrules}) at arbitrary values of
the reaction rates, specifically in the active phase, it is necessary to 
introduce a growth-limiting process. In most numerical simulations this is 
achieved by further imposing mutual exclusion between particles. Analytical 
progress therefore requires a consistent incorporation of the exclusion 
constraint. To this end, we follow Ref.~\cite{fredo}, and write down the 
resulting action corresponding to the processes (\ref{reactionrules}) on a 
(for the sake of notational simplicity one-dimensional) lattice (sites $i$):
\begin{eqnarray}
\label{lattaction}
  &&S[\{ {\hat \phi}_i , \phi_i \}] = \sum_i \int \! dt \ \Biggl( 
  {\hat \phi}_i \, \partial_t \phi_i + \biggl\{ 
  \mu \left( {\hat \phi}_i \, {\hat \phi}_{i+1} - 1 \right) \nonumber \\
  &&\qquad\qquad\qquad 
  + \frac{\mu'}{2} \left[ \left( {\hat \phi}_i - 1 \right) {\hat \phi}_{i+1} 
  + \left( {\hat \phi}_{i+1} - 1 \right) {\hat \phi}_i \right] \nonumber \\ 
  &&\, + \frac{\sigma}{2} \left[ \left( 1 - {\hat \phi}_{i-1} \right) 
  e^{- {\hat \phi}_{i-1} \phi_{i-1}} + \left( 1 - {\hat \phi}_{i+2} \right) 
  e^{- {\hat \phi}_{i+2} \phi_{i+2}} \right] \nonumber \\
  &&\qquad\qquad\quad 
  \times \ {\hat \phi}_i {\hat \phi}_{i+1} \biggr\} \ \phi_i \, \phi_{i+1} \ 
  e^{- {\hat \phi}_i \phi_i - {\hat \phi}_{i+1} \phi_{i+1}} \Biggr) \ .
\end{eqnarray}
The time-dependent fields ${\hat \phi}_i(t)$ and $\phi_i(t)$ here originate 
from a coherent-state representation employing {\em bosonic} creation and 
annihilation operators \cite{doietal,pelitilee}. The exclusion constraints are 
encoded in the exponential terms \cite{fredo}, and the unrestricted model is 
recovered when all these exponentials are replaced with unity.

Thus far, the action (\ref{lattaction}) constitutes an exact representation of
the microscopic processes (\ref{reactionrules}). In order to proceed to a
continuum field theory, which should suffice to describe the large-scale, 
long-time behavior in the vicinity of a critical point, we add a diffusion term
(for which we ignore the site occupation restrictions \cite{fnote2}) and take 
the (naive) continuum limit (now in $d$ dimensions) 
${\hat \phi}_i(t) \to {\hat \phi}(\mathbf{r},t)$ and 
$\phi_i(t) \to a^d \phi(\mathbf{r},t)$, with $a$ denoting the original lattice 
spacing, such that ${\hat \phi}(\mathbf{r},t)$ remains dimensionless. This 
yields
\begin{eqnarray}
\label{exclaction}
  &&S[{\hat \phi} , \phi] = \int \! d^dx \int \! dt \ \biggl\{ {\hat \phi} 
  \left( \partial_t - D \, \nabla^2 \right) \phi \nonumber \\
  &&\qquad\qquad\quad - \Bigl[ \mu \left( 1 - {\hat \phi}^2 \right) 
  + \mu' \left( 1 - {\hat \phi} \right) {\hat \phi} \Bigr] \phi^2 \ 
  e^{-2 v \, {\hat \phi} \phi} \nonumber \\
  &&\qquad\qquad\qquad + \sigma \left( 1 - {\hat \phi} \right) 
  {\hat \phi}^2 \, \phi^2 \ e^{-3 v \, {\hat \phi} \phi} \biggr\} \ ,
\end{eqnarray}
with a microscopic inverse density scale $v \sim a^d$.

The corresponding classical rate equation (augmented with diffusion) is readily
obtained by solving for the stationarity conditions 
$\delta S[{\hat \phi} , \phi] / \delta \phi = 0$, which, as a consequence of 
probability conservation in the master equation, is always satisfied by 
${\hat \phi} = 1$, and $\delta S[{\hat \phi} , \phi] / \delta {\hat \phi} = 0$,
which then results in
\begin{equation}
\label{mfrateq}
  \left( \partial_t - D \, \nabla^2 \right) \phi(\mathbf{r},t)
  = \phi^2 \left[ \sigma - \left( 2 \mu + \mu' \right) e^{v \, \phi} \right] 
  e^{-3 v \, \phi} \ .
\end{equation}
In contrast with the unrestricted model (where $v = 0$), this mean-field 
equation allows for an active state with a {\em finite} particle density 
$\phi = v^{-1} \ln (\sigma / \sigma_c)$, provided 
$\sigma > \sigma_c = 2 \mu + \mu'$. Near the now well-defined critical point at
$\sigma_c$, we obtain
\begin{equation}
\label{mfbeta}
  \phi(\sigma) \approx v^{-1} [(\sigma / \sigma_c) - 1]  
  \sim (\sigma - \sigma_c)^\beta \ , \quad \beta = 1 \ .
\end{equation}
In the absorbing phase ($\sigma < \sigma_c$), the site restrictions do not 
matter, and the density decays algebraically as in pure annihilation or
coagulation, $\phi(t) \sim [(\sigma_c - \sigma) t]^{-1}$. At the critical 
point, this becomes replaced with the slower power law
\begin{equation}
\label{mfdelta}
  \phi(t) \sim (v \sigma_c \, t)^{-1/2} \sim t^{-\delta }\ , \quad 
  \delta = 1/2 \ .
\end{equation}
This relation already shows that the scaling of the parameter $v$ determines
the critical exponents. Moreover, scaling analysis tells us that the static 
correlation length diverges upon approaching the critical point from the active
side according to $\xi(\sigma) \sim (\sigma - \sigma_c)^\nu$ with $\nu = 1$, 
whereas the characteristic time scales as $t_c \sim \xi^z$ with diffusive 
dynamic critical exponent $z = 2$.

While we clearly need to retain the exclusion parameter in order to describe 
the continuous phase transition occuring at $\Delta \propto \sigma - \sigma_c 
= 0$, we also note that an expansion to first order in $v$ suffices. More 
technically, since the field $\phi(\mathbf{r},t)$ scales as a particle density,
the scaling dimension of the exclusion parameter is $[v] = \kappa^{-d}$, where 
$\kappa$ represents an arbitrary momentum scale. Superficially, therefore, $v$ 
constitutes an {\em irrelevant} coupling that flows to zero under scale 
transformations. We may thus expand the exponentials in the action 
(\ref{exclaction}), keeping only the lowest-order contributions, which leads to
additional interaction terms. Upon at last performing the field shift 
${\hat \phi}(\mathbf{r},t) = 1 + \bar{\phi}(\mathbf{r},t)$ (whereby final-time
contributions, not explicitly listed here, become eliminated \cite{fnote3}), we
arrive at an action of the form
\begin{eqnarray}
\label{shifaction}
  &&S[\bar{\phi},\phi] = \int \! dx \int \! dt \ \Bigl[ \bar{\phi} \left( 
  \partial_t - D \, \nabla^2 \right) \phi + G(\bar{\phi}) \, \phi^2 \nonumber\\
  &&\qquad\qquad\qquad\qquad\qquad 
  + \Lambda(\bar{\phi}) \, \phi^3 + \ldots \Bigr] \ ,
\end{eqnarray}
where we have defined the {\em generating functions}
\begin{equation}
\label{functions}
  G(x) = \sum_{p \geq 1} g_p \, x^p \, , \quad 
  \Lambda(x) = \sum_{p \geq 1} \lambda_p \, x^p \ .
\end{equation}
Note that probability conservation implies $G(0) = 0$ and $\Lambda(0) = 0$.
Microscopically, we identify $g_1 = 2 \mu + \mu' - \sigma = \sigma_c - \sigma$,
$g_2 = \mu + \mu' - 2 \sigma$, 
$g_3 = - \sigma$, $\lambda_1 = (3 \sigma - 4 \mu - 2 \mu') v$, 
$\lambda_2 = (9 \sigma - 6 \mu - 4 \mu') v$, 
$\lambda_3 = (9 \sigma - 2 \mu - 2 \mu') v$, and $\lambda_4 = 3 \sigma \, v$. 
However, at a coarse-grained level, fluctuations generate the {\em entire} 
sequence of particle production reactions $2 A \to (n+2) A$ ($n \geq 1$). For 
instance, two subsequent branching processes $2 A \to 3 A$ immediately lead to 
$2 A \to 4 A$, and so forth \cite{martinuwe}. This effectively extends the sums
in the functions (\ref{functions}) to all integer $p$. Unlike in conventional 
situations, we thus have to deal with an {\em infinite} number of vertices.
Lastly we remark that introducing third-order annihilation reactions of the
form $3 A \to k A$ ($k = 0,1,2$) also produces the terms in the second line of
Eq.~(\ref{shifaction}): Allowing for the back reactions of the particle
production processes is equivalent to `soft' site exclusions.

With the previously introduced scaling dimensions of the fields 
$[\bar{\phi}] = \kappa^0$ and $[\phi] = \kappa^d$, we find 
$[g_p] = \kappa^{2-d}$ for all couplings in $G(x)$, which suggests, as is then
confirmed by a careful analysis of Feynman diagrams, that $d_c = 2$ constitutes
the upper critical dimension here. Since $[\lambda_p] = \kappa^{2 (1-d)}$, the 
coefficients in the function $\Lambda(x)$, which originates from site 
exclusion, are irrelevant near $d_c = 2$. Yet because at least $\lambda_1$ is 
required to control the particle density in the active phase and thereby 
maintain a continuous transition, it cannot simply be omitted from 
(\ref{shifaction}). Once again though we arrive at the conclusion that terms of
higher order in $v$ ({\em i.e.}, contributions $\sim \phi^4$ or higher in the 
action) need not be retained. But despite the presence of apparently infinitely
many marginal couplings, the field theory (\ref{shifaction}) remains 
renormalizable. This is best seen by recalling that the choice of the scaling 
dimensions for the fields is actually arbitrary as long as the product 
$[\bar{\phi} \, \phi] = \kappa^d$. Our theory thus contains a {\em redundant} 
variable \cite{wegner} that we fix conveniently as follows: Upon introducing 
recaled fields $\bar{s} = \kappa^{d/2} \bar{\phi}$ and 
$s = \kappa^{-d/2} \phi$, we obtain $[g_p] = \kappa^{2 - p \, d / 2}$ and
$[\lambda_p] = \kappa^{2 - (p+1) d / 2}$. Consequently, the critical control 
parameter constitutes a relevant perturbation (for $d < 4$), since
$[g_1] = \kappa^{2 - d / 2}$, whereas $[g_2] = [\lambda_1] = \kappa^{2 - d}$ 
indicating that both $g_2$ and $\lambda_1$ are marginal at $d_c = 2$. Indeed,
this procedure is consistent with the critical behavior according to 
Eq.~(\ref{mfdelta}) in two dimensions, which requires the density to scale 
$\sim t^{-1/2} \sim \kappa$ rather than $\sim t^{-1} \sim \kappa^2$ which is
valid in the absorbing phase. All other couplings now acquire negative scaling 
dimensions, and therefore become irrelevant for the leading scaling behavior. 
This leaves us with the reduced action
\begin{eqnarray}
\label{redaction}
  &&S[\bar{s},s] = \int \! dx \int \! dt \ \Bigl[ \bar{s} \left( 
  \partial_t - D \, \nabla^2 \right) s + g_1 \, \bar{s} \, s^2 \nonumber \\
  &&\qquad\qquad\qquad\qquad\qquad 
  + g_2 \, \bar{s}^2 s^2 + \lambda_1 \, \bar{s} \, s^3 \Bigr] \ .
\end{eqnarray}
Thus, the appropriate effective field theory for the critical point in fact 
only contains three nonlinear vertices.

\section{Renormalization and RG flow}
\label{rgflo}

It is instructive to proceed with the renormalization program based on either
the field theory (\ref{shifaction}) with infinitely many marginal couplings and
the reduced action (\ref{redaction}). One immediately notices that, to 
{\em all} orders in the perturbation expansion, the propagators do not become
renormalized. Hence there is neither field nor diffusion constant 
renormalization, which already implies that the dynamic exponent in these field
theories inevitably remains $z = 2$ {\em exactly}, at variance with present 
simulation data. Next, for the action (\ref{shifaction}) we define renormalized
parameters according to 
$g_p \, C_\epsilon = Z_p \, g_{p {\rm R}} \, D \, \kappa^\epsilon$, and 
$\lambda_p \, C_\epsilon = Z_{\lambda_p} \, \lambda_{p {\rm R}} \, D \, 
\kappa^{-2 + 2 \epsilon}$ with 
$C_\epsilon = (4 \pi)^{- d/2} \, \Gamma(1 + \epsilon / 2)$. The renormalization
constants $Z_p$ and $Z_{\lambda_p}$ are determined by means of dimensional 
regularization and minimal subtraction by the condition that they absorb just 
the ultraviolet divergences appearing as poles in $\epsilon = 2 - d$. After 
computing the RG $\beta$ functions 
$\beta_p = \kappa \, \partial_\kappa g_{p {\rm R}}$ (evaluated in the
unrenormalized theory) and upon introducing the flow parameter 
$\ell = - \ln (\kappa a)$ (where $\kappa$ is the running momentum scale), we 
subsequently obtain the corresponding RG flow equations for the running 
couplings, $\partial_\ell \, g_{p {\rm R}}(\ell) = - \beta_p(\ell)$, and 
similarly for the $\lambda_{p {\rm R}}$.

We start with renormalizing the action (\ref{redaction}). The vertex
$\sim \lambda_1$ does not enter any Feynman diagrams that contribute to the
renormalization of $g_1 \, C_\epsilon = Z_1 \, g_{1 {\rm R}} \, D \, 
\kappa^{1 + \epsilon / 2}$ and 
$g_2 \, C_\epsilon = Z_2 \, g_{2 {\rm R}} \, D \, \kappa^\epsilon$. For the 
latter, we are therefore left with precisely the structure of the pure 
annihilation / coagulation field theory \cite{pelitilee}, namely a geometric
series of one-loop graphs, and hence arrive at the {\em exact} result
\begin{equation}
\label{exactz}
  Z_1^{-1} = Z_2^{-1} = 1 - \frac{2 \, g_{2 {\rm R}}}{\epsilon} \ . 
\end{equation}
This leaves $Z_{\lambda_1}$ in $\lambda_1 \, C_\epsilon = 
Z_{\lambda_1} \, \lambda_{1 {\rm R}} \, D \, \kappa^\epsilon$ as the sole 
renormalization constant to be actually determined anew here. To two-loop 
order, we find \cite{fnote4}
\begin{equation}
\label{renconst}
  Z_{\lambda_1}^{-1} = 1 - \frac{6 \, g_{2 {\rm R}}}{\epsilon} 
  + \frac{12 \, g_{2{\rm R}}^2}{\epsilon^2} 
  \left( 1 + \frac{\epsilon}{2} \, \ln \frac 43 \right) \, .
\end{equation}

\begin{figure}[ptb]
\includegraphics*[scale=0.45,angle=0]{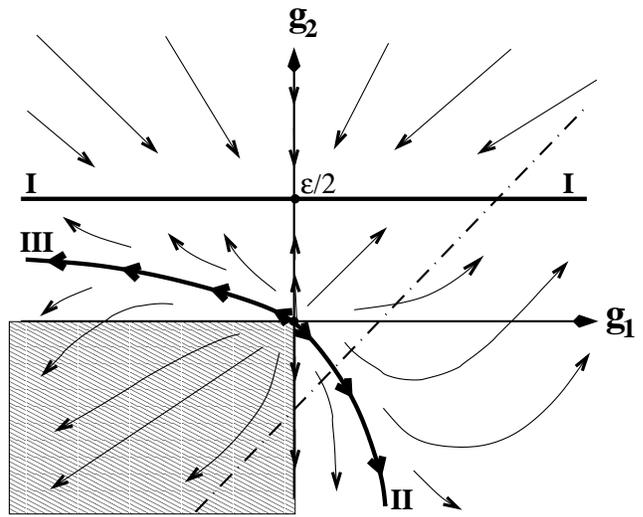}
\caption{Qualitative flow diagram of the RG trajectories in the 
  ($g_1,g_2$)-plane. {\textbf{I}} denotes the stable fixed line; {\textbf{II}} 
  and {\textbf{III}} represent the parts of the separatrix that might collapse 
  in the limiting case with the negative half-axes. The hatched area indicates 
  the absolutely unstable region. The dash-dotted line corresponds to the 
  initial values of the model, with $2 g_1 - g_2 = {\rm const.} > 0$.
  \label{flofig}}
\end{figure}
From Eqs.~(\ref{exactz}) we obtain the RG flow equations
\begin{equation}
\label{g12beta}
  \partial_\ell \, g_{1/2 {\rm R}}(\ell) = \bigl[ \epsilon 
  - 2 \, g_{2 {\rm R}}(\ell) \bigr] \, g_{1/2 {\rm R}}(\ell) \ .
\end{equation}
It follows that the sign of $g_{1 {\rm R}}$ is invariant under the RG flow, and
the critical $g_{1 {\rm R}} = 0$ remains fixed. If both $g_{1 {\rm R}}$ and 
$g_{2 {\rm R}}$ are negative, the flow according to Eqs.~(\ref{g12beta}) leads 
both running couplings towards $- \infty$  (`Wilson's gully'). Only if either
parameter is initially positive, can the stable fixed line with arbitrary 
$g_{1 {\rm R}}$ and $g_{2 {\rm R}}^* = \epsilon / 2$ be reached. Thus, a
separatrix must exist between the fully unstable region in parameter space and
the basin of attraction of the fixed line. A qualitative sketch of the ensuing 
RG trajectories is depicted in Fig.~\ref{flofig}.

Before we proceed further with a discussion of the RG flow trajectories, let us
consider the field theory (\ref{shifaction}), and for the moment omit the 
irrelevant parameters $\lambda_p$. One is then concerned with controlling the 
infinite number of marginally relevant vertices $g_p$. This is most elegantly 
achieved by introducing the renormalized counterpart of the generating function
$G_{\rm R}(x) = \sum_{p \geq 1} g_{p {\rm R}} \, x^p$ \cite{john}. As 
anticipated, a careful analysis of the appropriate Feynman diagrams indeed 
shows that not only are the renormalizations of the $g_p$ interwined, but also 
such couplings of arbitrarily high order become generated. For example, an 
explicit two-loop calculation results in 
\begin{eqnarray}
\label{facteurZ-g_2}
  &&\!Z_2^{-1}\! = 
  1 - \frac{2 g_2 \, C_\epsilon \kappa^{-\epsilon}}{D \, \epsilon} - 
  \frac{6 g_1 g_3 \, C_\epsilon \kappa^{-\epsilon}}{D g_2 \, \epsilon} + 
  \frac{24 g_1^2 g_4 C_\epsilon^2 \kappa^{-2 \epsilon}}{D^2\, g_2\, \epsilon^2}
  \nonumber \\ 
  &&\ + \frac{4 g_2^2 \, C_\epsilon^2 \kappa^{-2 \epsilon}}{D^2 \, \epsilon^2}
  + \frac{12 g_1 g_3 \, C_\epsilon^2 \kappa^{-2 \epsilon}}{D^2 \, \epsilon^2} 
  \left( 1 + \frac{\epsilon}{2} \, \ln \frac43 \right) .
\end{eqnarray}
To one-loop order, elementary combinatorics yields \cite{martinuwe}
\begin{equation}
\label{zfactors}
  Z_p^{-1} = 1 - \frac{1}{\epsilon} \, \sum_{j=1}^p j (j+1) \ 
  \frac{g_{j+1 {\rm R}} \, g_{p-j+1 {\rm R}}}{g_{p {\rm R}}} \ ,
\end{equation}
whence $\beta_p = - \epsilon \, g_{p {\rm R}} + \sum_{j=1}^p j (j+1) \,
g_{j+1 {\rm R}} \, g_{p-j+1 {\rm R}}$. The ensuing infinite hierarchy of 
ome-loop flow equations for the $g_{p {\rm R}}$ is then efficiently recast into
a {\em single} functional RG differential equation for $G_{\rm R}(x,\ell)$ 
\cite{john}:
\begin{equation}
\label{jolie-equation}
  \partial_\ell \, G_{\rm R}(x,\ell) = \left[ \epsilon - 
  \partial_x^2 \, G_{\rm R}(x,\ell) \right] G_{\rm R}(x,\ell) \ .
\end{equation}
Although we shall not explicitly make use of it, one may derive in a similar 
fashion the one-loop functional RG flow equation for the generating function 
$\Lambda_{\rm R}$ that incorporates all the couplings induced by the site 
exclusions:
\begin{eqnarray}
\label{rglambda}
  &&\partial_\ell \, \Lambda_{\rm R}(x,\ell) = -2 (1 - \epsilon) \, 
  \Lambda_{\rm R}(x,\ell) \\ 
  &&\ - G_{\rm R}(x,\ell) \ \partial_x^2 \, \Lambda_{\rm R}(x,\ell) - 3 \, 
  \Lambda_{\rm R}(x,\ell) \ \partial_x^2 \, G_{\rm R}(x,\ell) \nonumber \ .
\end{eqnarray}

Recall that the initial generating function was a third-order polynomial 
$G_{0 {\rm R}}(x) = g_{1 {\rm R}} \, x + g_{2 {\rm R}} \, x^2 + 
g_{3 {\rm R}} \, x^3$ with at least $g_{3 {\rm R}} < 0$. Upon setting the 
right-hand side of Eq.~(\ref{jolie-equation}) to zero, we find the locally 
stable nontrivial fixed-point function
\begin{equation}
\label{fpointf}
  G_{\rm R}^*(x) = \Delta \, x + \frac{\epsilon}{2} \, x^2
\end{equation}
with arbitrary constant $\Delta$, whereas the trivial solution 
$G_{\rm R}^* = 0$ is clearly unstable. This result is in fact valid to 
{\em all} orders in $\epsilon$ \cite{fnote5}. Therefore, all running couplings 
$g_{p {\rm R}}(\ell) \to 0$ for $p \geq 3$, and we recover the results 
(\ref{exactz}), (\ref{renconst}), and (\ref{g12beta}) previously obtained from 
the reduced action (\ref{redaction}). As anticipated, $\Delta$, the 
renormalized counterpart to $g_1$, plays the role of a control parameter, 
albeit not a relevant one, since that would have to scale to infinity under
renormalization rather than remain constant. We assume that $\Delta$ is a 
regular function of the initial rates; {\em i.e.}, at fixed annihilation and 
coagulation rates we expand for $\sigma \to \sigma_c$: 
$\Delta(\sigma) \simeq (\sigma_c - \sigma) \, \Delta'(\sigma_c)$, since 
$\Delta(\sigma_c) = 0$ and $\Delta'(\sigma_c) > 0$. Indeed, the pure 
annihilation and coagulation model fixed points at $\sigma = 0$ respectively 
correspond to $\Delta = \epsilon$ (since $g_1 = 2 g_2$) and 
$\Delta = \epsilon / 2$ ($g_1 = g_2$) \cite{pelitilee}. Thus, in the PCPD 
inactive phase $\sigma < \sigma_c$ one should have 
$\Delta = {\cal O}(\epsilon) > 0$ as well. Notice that at 
$\sigma_c = 2 \mu + \mu'$ we have $g_2 = -3 \mu - \mu'$, hence this combination
of annihilation and coagulation rates has apparently turned negative at the 
fixed point. One must therefore worry whether the physically accessible values 
of the reaction rates actually lie within the basin of attraction of the
nontrivial fixed-point function $G_{\rm R}^*(x)$. 

Thus we now resume our discussion of the RG trajectories. On physical grounds 
one should expect that the flow would at least reach the line of fixed points 
encoded by Eq.~(\ref{fpointf}) if $g_{1 {\rm R}}$ is positive, since this 
corresponds to the inactive phase governed by the annihilation / coagulation 
fixed line. Yet for this to be true for any $g_{1 {\rm R}} > 0$, the part of 
the separatrix indicated by II in Fig.~\ref{flofig} must collapse onto the 
negative $g_2$-axis, {\em i.e.}, the separatrix should contain the invariant 
hypersurface $g_1 = 0$. This would indeed lead to the standard RG flow picture:
With initial values corresponding to the inactive phase the RG trajectories
approach the annihilation / coagulation fixed line, critical initial conditions
are defined by the separatrix (which is unstable along one direction), and
finally initial values corresponding to the active phase are to be found inside
the unstable region with flow into the `gully'.
 
Yet it is easily seen that the zeros $x_i$ with $G_{\rm R}(x_i,\ell) = 0$ are 
fixed by the partial differential equation (\ref{jolie-equation}), and 
consequently also the intervals in which the function $G_{\rm R}(x,\ell)$ is 
respectively positive or negative. Therefore, if (i) $g_{1 {\rm R}} > 0$, there
exists an open interval $(x_0 = 0, x_1 > 0)$ wherein $G_{\rm R}(x,\ell) > 0$, 
while $G_{\rm R}(x_{0,1},\ell) = 0$. Since initially $g_{3 {\rm R}} < 0$, 
however, it follows that the fixed-point function (\ref{fpointf}) {\em cannot} 
be reached from the physically allowed region in parameter space. This is true
for arbitrary values of $\Delta$ and for all $-\infty < x < \infty$, even if 
$g_{2 \rm R} > 0$. This is rather astonishing because one would expect that at 
least deep in the inactive phase, where the production reactions can be 
neglected, $G_{\rm R}(x,\ell) \to G_\infty(x) = G_{\rm R}^*(x)$ as 
$\ell \to \infty$ for all $x$. It is however obvious that we can assume this 
limiting relation to hold only in the first positive interval in an expansion 
of $G_{\rm R}(x)$ with respect to $x$. We conclude that the difference function
$H(x) = G_{\rm R}^*(x) - G_\infty(x)$ must display an essential singularity at 
$x = 0$, and its expansion in a series of $x$ produces simply a zero. However,
$H(x_1) = G_{\rm R}^*(x_1)$. A qualitative discussion of the differential
equation (\ref{jolie-equation}) indeed supports this assumption. Note here that
$G_{\rm R}(x,\ell)$ increases most significantly with $\ell$ at those values of
$x$ where its curvature is minimal. (ii) In the case $g_{1 {\rm R}} \leq 0$ the
initial function is negative for all $x > 0$. Hence the differential equation
(\ref{jolie-equation}) does not provide any mechanism that could translate 
$G_{\rm R}(x,\ell)$ to positive values and finally to $G_{\rm R}^*(x)$, at 
least in some $x$-intervals, without producing zeros of $G_{\rm R}(x)$ along 
the way. In summary, we find that our physical initial conditions at the
critical point (with $g_{3 {\rm R}} < 0$ as well as $g_{2 {\rm R}} < 0$ for 
$g_{1 {\rm R}} = \Delta = 0$) inevitably take the RG flow from 
Eq.~(\ref{jolie-equation}) into the absolutely unstable region in 
Fig.~\ref{flofig}. Instead of reaching the stable fixed line (\ref{fpointf}),
we face runaway trajectories, to {\em all} orders in the perturbation 
expansion.

\section{Critical properties at the unphysical fixed point}
\label{crexp}

Even though we have just seen that the critical fixed point 
$g_{1 {\rm R}} = \Delta = 0$ and $g_{2 {\rm R}}^* = \epsilon / 2$ is 
inaccessible to the RG flow trajectories starting at physical initial parameter
values ({\em i.e.}, positive reaction rates), let us nevertheless explore the
(hypothetical) ensuing critical behavior. Recall that in the PCPD, totally 
neglecting particle exclusion, as encoded in the parameter $\lambda_1$, 
suppresses the finite-density steady-state. Hence we retain this (apparently
irrelevant) coupling, and moreover investigate how its RG flow towards zero 
becomes renormalized through fluctuations. From the explicit two-loop result 
for the associated renormalization constant (\ref{renconst}), we may 
immediately compute the anomalous dimension
\begin{equation}
\label{anomdim}
   \gamma_{\lambda_1} = \kappa \, \frac{d \ln Z_{\lambda_1}}{d \kappa} =
   - 3 \, \epsilon + 3 \, \epsilon^2 \, \ln \frac 43 + {\cal O}(\epsilon^3) \ .
\end{equation}

As is easily seen by investigating the RG equations for the particle density
and its correlation function, $\gamma_{\lambda_1}$ already completely 
determines the critical exponents here. This assertion is also confirmed by
directly computing the renormalized equation of state (upon approaching the 
transition from the active side). For $\Delta \sim \sigma - \sigma_c \geq 0$, 
{\em i.e.} in the active phase, one finds that the steady-state density 
vanishes as $\Delta \to 0^-$ according to 
$\langle \phi \rangle \sim |\Delta|^\beta$, with
\begin{eqnarray}
\label{beta}
  &&\beta = \frac{1 + \gamma_{\lambda_1} / 2}{d - 1 -\gamma_{\lambda_1} / 2} 
  \nonumber \\
  &&\quad \simeq 1 - 2 \, \epsilon + \left( 1 + 3 \ln\frac 43 \right) 
  \epsilon^2 + {\cal O}(\epsilon^3) \ ,
\end{eqnarray}
while the two-point function correlation length (finite only in the active 
phase) diverges as $\xi \sim |\Delta|^{-\nu}$, where
\begin{equation}
\label{invnu}
  \nu^{-1} = d - 1 - \frac{\gamma_{\lambda_1}}{2} \simeq 1 + \frac{\epsilon}{2}
  - \frac 32 \, \epsilon^2 \, \ln\frac 43 + {\cal O}(\epsilon^3) \ .
\end{equation}
Precisely at the critical point ($\Delta = 0$) the particle density decays 
asymptotically as $\langle \phi(t) \rangle \sim t^{-\delta}$, with
\begin{equation}
\label{delta}
  \delta = \frac 12 + \frac{\gamma_{\lambda_1}}{4} \simeq \frac 12 - \frac 34\,
  \epsilon + \frac 34 \, \epsilon^2 \, \ln \frac 43 + {\cal O}(\epsilon^3) \ .
\end{equation}

Since remarkably the anomalous fluctuation conrrections to the critical 
exponents $\beta$, $\nu$, and $\delta$ are {\em solely} contained in the 
anomalous dimension (\ref{anomdim}), we may eliminate the latter to yield the 
following {\em hyperscaling relations}, valid to {\em all} orders in 
$\epsilon = 2 - d$:
\begin{equation}
\label{scarel}
  2 \, \delta + 1/\nu = d \ , \quad 
  \beta = 2 \delta / (d - 2 \delta) = d \, \nu - 1 \ .
\end{equation}
Here we have used the exact result for the dynamic exponent $z = 2$ and the 
standard scaling relation $\beta = z \, \nu \, \delta$ (which also follows 
directly from the RG equations). At the critical dimension $d_c = 2$, we infer 
the asymptotically exact scaling behavior
\begin{eqnarray}
  &\Delta \to 0^-: \ &\langle \phi \rangle \sim |\Delta| \, (\ln |\Delta|)^2 
  \ , \\ &&\ \xi \sim |\Delta|^{-1} \, \big| \ln |\Delta| \big|^{-1/2} \ , \\
  &\Delta = 0: \ &\langle \phi(t) \rangle \sim t^{-1/2} \, (\ln t)^{3/2} \ .
\end{eqnarray}

Aside from the fact that these exponent values are at odds with the presently
available data from numerical simulations for the PCPD, they also lead to a 
serious contradiction, which confirms again that the fixed line (\ref{fpointf})
does not represent a physical system. First, we note that the positive value
$g_{2 {\rm R}}^* = \epsilon / 2$ indicates the presence of particle
{\em anti}correlations at this fixed point, precisely as in the pure binary
annihilation or coagulation system. Next, recall that the equal-time density 
correlation function of point-like particles consists of three contributions,
\begin{eqnarray}
\label{Funda}
  &&\langle n(\mathbf{r},t) \, n(\mathbf{r}',t) \rangle = 
  \langle n(\mathbf{r},t) \rangle \, \delta(\mathbf{r} - \mathbf{r}') 
  \nonumber \\ &&\quad\qquad 
  + C(\mathbf{r},\mathbf{r}';t) + \langle n(\mathbf{r},t) \rangle \, 
  \langle n(\mathbf{r}',t) \rangle \ .
\end{eqnarray}
The first term here describes the particles' Poissonian self-correlations. 
$C(\mathbf{r},\mathbf{r}';t)$ represents the density cumulant, {\em i.e.}, the 
connected correlation function of the density fluctuations, which is negative 
in the case of particle anticorrelations. Upon integrating Eq.~(\ref{Funda}) 
over the confining volume and dividing by the mean particle number 
$\langle N(t) \rangle$, we obtain for a homogeneous state where 
$C(\mathbf{r},\mathbf{r}';t) = C(\mathbf{r} - \mathbf{r}';t)$ the following
general expression for the relative mean-square particle number fluctuations:
\begin{equation}
\label{FlukForm}
  \frac{\langle \delta N(t)^2 \rangle}{\langle N(t) \rangle}
  = 1 + \frac{{\widetilde C}(t)}{\phi(t)} \ , 
\end{equation}
with the mean density $\phi(t) = \langle n(\mathbf{r},t) \rangle$ and the 
spatial integral of the cumulant ${\widetilde C}(t)$. 

In the vicinity of a critical point these quantities scale as follows:
\begin{equation}
\label{AsSkal}
  \phi(t) \sim A \, t^{- \beta / z \nu} \ , \quad 
  {\widetilde C}(t) \sim B \, t^ {(d \nu - 2 \beta) / z \nu} \ .
\end{equation}
The amplitude $A$ is of course positive, while $B < 0$ in the case of 
anticorrelations (such as in the PC universality class), and $B > 0$ for 
positive particle correlations (as prevalent in the critical DP clusters).
Combining Eqs.~(\ref{FlukForm}) and (\ref{AsSkal}) yields
\begin{equation}
\label{FlukAs}
  \frac{\langle \delta N(t)^2 \rangle}{\langle N(t) \rangle} \sim
  1 + \frac{B}{A} \ t^{(d \nu - \beta) / z \nu} \ .
\end{equation}
Thus, if $d \nu - \beta > 0$ the second term dominates the right-hand side of 
Eq.~(\ref{FlukAs}) asymptotically. For particle anticorrelations where 
$B / A < 0$, this would immediately contradict the positivity of the 
left-hand side. Consequently, the previously found scaling exponents which
satisfy $d \nu - \beta = 1$ (exactly) are definitely inacceptable.

This is in remarkable contrast to the results obtained for the pure pair
annihilation or coagulation process \cite{pelitilee}, where the density and the
integrated cumulant obey identical asymptotic scaling behavior, 
$\phi(t) \sim {\widetilde C}(t) \sim t^{-d/2}$. Hence no contradiction arises 
here provided $|B| < A$. Neither are particle anticorrelations necessarily
excluded at critical points: For even-offspring BAW that represent the PC
universality class, a one-loop RG analysis at fixed dimension yields
$\nu = 3 / (10 - 3 d)$ and $\beta = 4 / (10 - 3 d)$ \cite{johnuwe}. Hence
$d \nu - \beta \leq 0$ in dimensions $d \leq d_c' = 4/3$, which is precisely 
the borderline dimension (within the one-loop approximation) for the existence 
of the phase transition and the power-law inactive phase in this system.

\section{Conclusions}
\label{concl}

We have investigated the `microscopic' field theory for the PCPD, as derived
directly from the defining master equation, by means of the dynamic 
renormalization group. Although fluctuations generate an infinite chain of 
particle production processes $A + A \to (n+2) \, A$, the theory remains 
controlled and renormalizable in the inactive phase \cite{martinuwe}, where it 
is governed by the fixed point of the pure annihilation / coagulation model 
\cite{pelitilee}. This is most elegantly seen by means of a functional RG 
approach \cite{john}. In order to render the particle density finite in the 
active phase, we have incorporated site occupation restrictions following the 
methods developed in Ref.~\cite{fredo}. On the mean-field level, this indeed 
leads to a continuous transition separating the active from the absorbing 
phase. However, a detailed analysis of the RG flow equations shows that the 
action (\ref{shifaction}) does {\em not} adequately capture the critical 
properties of the PCPD: (i) There is no stable RG fixed point in the physical 
region of parameter space , and instead one obtains runaway trajectories; (ii) 
the scaling exponents found at the RG fixed point in the unphysical regime 
violate the positivity of the mean-square particle number fluctuations. 

We emphasize that these statements in fact hold to {\em all} orders of the 
perturbation expansion and even apply to the nonperturbative `exact' RG 
approach. This failure really resides in the starting field theory action 
itself, not its subsequent evaluation. Obviously, a crucial ingredient was left
out when the `naive' continuum limit was performed. We can only speculate as to
what the potential remedies might be, in part motivated by pictures from 
simulation studies, where positive particle correlations are observed both at 
the critical point and in the active phase: Perhaps one needs to explicitly 
introduce separate fields respectively for the positively correlated clustered 
particles and the solitary random walkers as coupled slow variables. The 
challenge then, however, is to write down a consistent coarse-grained field 
theory that correctly accounts for the internal stochastic noise generated by 
the reactions. To date, therefore, an apt field theory description of the PCPD 
remains an open and difficult problem.

Finally, we remark that the same problems arise with the master equation field
theory when the PCPD order parameter is coupled to a static, conserved 
background field \cite{kockelkorenchate2}. The above statements also apply to 
active-to-absorbing transitions in related higher-order reactions 
\cite{parkhinrichsenkim2,kockelkorenchate1,odor4}. For purely triplet 
reactions, $3 A \to m A$ ($m = 0,1,2$) and $3 A \to (n+3) \, A$ one readily 
derives a field theory representation corresponding to the action 
(\ref{shifaction}), essentially just raising the powers of $\phi$ by one in the
nonlinear vertices. In the inactive phase, the theory can again be analyzed by 
means of the functional RG analogous to Eq.~(\ref{jolie-equation}), leading to 
the upper critical dimension $d_c = 1$, where the particle density decays 
according to $\langle \phi(t) \rangle \sim (t^{-1} \ln t)^{1/2}$ 
\cite{pelitilee}. At the hypothetical critical point one would obtain a slower 
critical decay $\langle \phi(t) \rangle \sim t^{-1/3} \, (\ln t)^{4/3}$. But 
once again, the critical fixed point cannot be reached by the RG flow starting
at physical parameter values. Moreover, the presence of a redundant parameter
even questions the validity of identifying $d = 1$ as the critical dimension.
Likewise, the expectation that in general fourth-order processes are merely 
described by mean-field scaling exponents may not be borne out by a correct
treatment.

\begin{acknowledgments}
FvW thanks Fundamenteel Onderzoek der Materie and the Lorentz Fonds for 
financial support while part of this work was carried out.
UCT is grateful for support by the National Science Foundation (NSF award 
DMR-0308548), and by the Jeffress Memorial Trust (grant no. J-594). 
We gladly acknowledge fruitful discussions with G. Barkema, H. van  Beijeren, 
J. Cardy, E. Carlon, H. Chat\'e, H.W. Diehl, I. Dornic, P. Grassberger, 
M. Henkel, H. Hinrichsen, M. Howard, M.-A. Mu\~noz, M. den Nijs, G. \'Odor, 
B. Schmittmann, and G. Sch\"utz.
\end{acknowledgments}


\begin{thebibliography}{}
\bibitem{fnote1}
  The equilibrium theory of classical antiferromagnets might provide a useful 
  analogy here: A `naive' continuum limit of the Ising or Heisenberg models 
  directly leads to an action that contains only the local magnetization. In 
  order to describe the physics correctly in terms of an appropriate effective
  coarse-grained theory, one must first introduce the staggered magnetization 
  as the suitable order parameter, and only then carefully proceed with the 
  continuum limit.
\bibitem{fnote4}
  Remarkably, an identical two-loop result is found for the anomalous dimension
  of the branching rate in BAW with $m = 2$ \cite{johnuwe}, albeit based on 
  Feynman diagrams with inverted time arrows.
\end{thebibliography}
\end{document}